# TALLMesh: a simple application for performing Thematic Analysis with Large Language Models

Stefano De Paoli and Alex Fawzi

Abertay University, Dundee

s.depaoli@abertay.ac.uk

a.fawzi@abertay.ac.uk

[Draft, please get in touch with the authors if spot inconsistencies or if you have questions]

## Abstract

Thematic analysis (TA) is a widely used qualitative research method for identifying and interpreting patterns within textual data, such as qualitative interviews. Recent research has shown that it is possible to satisfactorily perform TA using Large Language Models (LLMs). This paper presents a novel application using LLMs to assist researchers in conducting TA. The application enables users to upload textual data, generate initial codes and themes. All of this is possible through a simple Graphical User Interface, (GUI) based on the streamlit framework, working with python scripts for the analysis, and using Application Program Interfaces of LLMs. Having a GUI is particularly important for researchers in fields where coding skills may not be prevalent, such as social sciences or humanities. With the app, users can iteratively refine codes and themes adopting a human-in-the-loop process, without the need to work with programming and scripting. The paper describes the application's key features, highlighting its potential for qualitative research while preserving methodological rigor. The paper discusses the design and interface of the app and outlines future directions for this work.

## Introduction

Large Language Models are generative Artificial Intelligence (AI) systems that manipulate language, and amongst others, they excel at tasks such as summarisation or classification of texts. LLMs can achieve this because they are trained on large corpuses of textual data. Recent solutions like the OpenAI chatGPT or Anthropic Claude, are known examples of chatbot applications that use LLMs. However, LLMs can also function as standalone solutions, useful for building independent applications to perform defined and specific tasks, especially (but not only) if used through an Application Program Interface (API).

Within the field of social sciences, one of the most exciting practical applications of LLMs has been the methodological research conducted for performing different types of qualitative data analysis. Qualitative analysis, differently from quantitative analysis, is a subjective approach where the analysts seek to extract meaning from rich textual data such as qualitative interviews, online posts, or news articles for example. The qualitative analysis approaches are many and differing, and include for example Grounded Theory, Thematic Analysis, Phenomenological Analysis or Discourse Analysis. Often, however, most of the qualitative analysis procedures are based on some common or similar activities and differ mostly in their objectives as well as in some components of the interpretation process. For example, the initial coding of data is common across a wide variety of methods. The initial coding entails the analysts labelling portions of the text, with labels that seek to capture the explicit or latent meaning of the text. This is an analysis phase where large textual data is broken down into small and more manageable parts associated with relevant meaning and useful for answering the research question (Saldaña, 2021).

Thematic analysis (TA) has been singled out in various literature contributions, as a qualitative approach that could be reproduced with the support of LLMs (see e.g. De Paoli, 2024a). TA was famously systematised and popularised by Braun and Clarke (2006) who stipulated that this method encompasses six key phases: (1) familiarisation with the data; (2) initial coding; (3) formulation of themes; (4) revision of themes; (5) renaming of themes; (6) write up of the results. Familiarisation with the data entails the analysts reading carefully the material (e.g. interview transcripts) to get an idea about the content. Themes are patterns in the dataset and are the main expected results of this analysis approach, and they are created by sorting and grouping initial codes together. The revision and renaming of themes are steps for further refining the themes and improving the analysis results, followed by their write up of the results.

Initial coding and generation of themes, as critical steps for a TA, appeared possible to be reproduced using LLMs, with appropriate prompting through a programmatic approach using LLMs APIs. Some details of these steps will be briefly discussed later. This normally entails using scripts, for example in python language, which load interviews into a prompt (the instructions a user provides to the LLM to perform a defined task). Whilst the overhead for performing these steps with programming is not enormous, it still requires some mastery of coding, prompting, the use of remote solutions through APIs or endpoints, or various software libraries. Researchers and analysts with limited or no programming skills will find this difficult. This is particularly relevant for researchers in the broader field of social sciences and humanities, where coding skills are not widespread (e.g. Sufi et al., 2021; Harold et al., 2023). To facilitate the use of the method via LLMs, having a graphical user interface (GUI) would facilitate adoption, especially for researchers who do not program. This paper's objective is to describe the app that resulted from work we conducted to create a GUI, for the python script we previously defined in our research (see De Paoli, 2024a and 2024b; De Paoli and Mathis, 2024, Mathis et al., 2024) for performing TA. We call the app TALLMesh (Thematic Analysis with LLMs). We will describe the key features of the app through its interface, and the work we performed to improve its usability. The app is available as a beta open-source software solution from GitHub (https://github.com/sdptn/TALLMesh_multi_page).

# Literature and apps review

This section briefly discusses the current state-of-the-art by reviewing selected literature on qualitative analysis using LLMs, particularly thematic analysis (TA), before examining several existing apps and graphical user interfaces designed to support researchers performing this type of analysis.

## Literature

A TA with LLMs, in various forms, reproduces most of the phases defined by Braun and Clarke (2006), with some differing levels of innovation to accommodate some of the limits of LLMs. Without claiming superiority over existing approaches, we describe here some elements of our own procedure, concentrating on the steps of initial coding and the definition of themes. This is because later this material will facilitate understanding the interface of TALLMesh. For initial coding, interviews are passed to an LLMs for example using the Application Program Interface, through python script and a prompt which normally instructs the LLM to extract initial codes, and quotes from interviews, alongside producing short description of each code (see Figure 1 for a description of this step and De Paoli 2024a for the entire process and prompts). Following this, the list of initial codes can then be passed again to the LLM programmatically, with a prompt instructing the model to sort and aggregate the codes into themes. These are in essence two of the key steps of the process, albeit underneath these, methodological research was conducted to establish reproducible procedures as well as evaluating quality and validity.

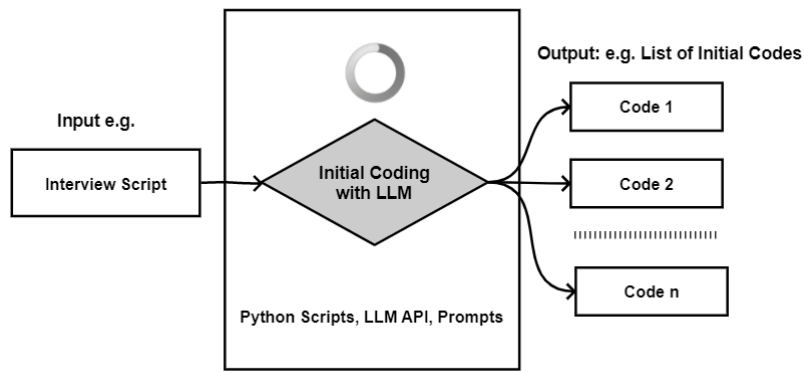

Figure 1. Initial coding of interviews with LLMs, programming scripts and an API

Amongst other, seminal works on performing TA with LLMs programmatically was performed around the late 2023/early 2024 by a few authors including De Paoli (2024a), Gao et al. (2023), or Xiao et al. (2023), with the establishment of some initial, perhaps still crude procedures, and in some cases the definition of working prompts that could achieve some of the steps of TA. Some papers only focused on initial coding, whilst others attempted the entire process of TA (see e.g. De Paoli, 2024b). Some authors used directly the chatbot versions of LLMs (e.g. chatGPT), for instance Kon et al. (2024) or Jalali and Akhavan (2024). However, these procedures using the chatbots offer, in our opinion, less scalability and data processing power compared to using programming scripts. After a first initial (small) wave of contributions which proved the feasibility of using LLMs for TA and qualitative analysis, procedures were then refined and improved in further contributions such as De Paoli and Mathis (2024), Liu et al. (2024), Zhao et al. (2024), Törnberg (2024) or Yang and Ma (2025) amongst others.

Alongside establishing the methodological procedures, following contributions have also then concentrated on better assessing the quality of the methods. This includes normally establishing evaluation procedures, such as using multiple LLMs for comparisons (e.g. Drapal et al., 2024), using sentence transformers and/or human assessments of LLM outputs (e.g. Mathis et al., 2024; Deiner et al. 2024), redefining existing quality procedures such as saturation of initial codes to work for LLMs (e.g. De Paoli and Mathis, 2024) or developing novel computational method to assess open coding and reduce 'coding bias' (Chen et al. 2024).

## Review of Some Existing apps for TA and Initial Coding with LLMs

Except for the solutions using directly chatbots, performing TA (or qualitative analysis) with LLMs requires some mastery of programming, which may not always be immediately available as a skill to social scientists, for example. This is probably the reason why few authors have performed research to design solutions which offer researchers graphical user interfaces, which require minimal or no coding skills. We will review some of these applications briefly.

Gao et al. (2024) proposed *Collabcoder*. Their aim was to design a one-stop, end-to-end workflow, to lower the barrier for performing qualitative coding. Moreover, the objective of this solution is to support collaboration and agreement between multiple coders (as the name of the solution implies). The objective then appears to bring capacity to deliver initial qualitative coding to researchers which might have limited knowledge of qualitative analysis. The interface which can be seen in an available demo[1] offers essentially three steps: 1) generation of codes with LLMs and selection from codes suggestions, 2) comparison of similarity of choices between two coders and 3) creation of a final codebook. The existing interface does also offer a measure of agreement between the coders as well as the measure of intercoder reliability (with a calculation of a metric called Cohen's K).

---

[1] See http://16.16.202.78/Demouser/demo/team/Demouser

*Qualcoder*[2], proposed by Curtin (2023), is an Open-Source Qualitative Data Analysis Software (QDAS) solution which has been around for few years and has already a substantial and developing user base. The software is designed for manual qualitative coding and analysis, and it supports different data types including text, audio or video. As this was originally developed as a standard QDAS the interface does present similarity with more traditional and commercial solutions as expected. Qualcoder has been recently integrated[3] with some AI features including an AI coding assistant and AI chat, supporting commercial models such as GPT4 and Open-Source models.

*LLMCode*[4] (Hämäläinen, et al., 2024) is a follow-on app from a paper by Hämäläinen et al. (2023) on using LLMs for generating synthetic data for HCI research. The GUI of the app is a notebook interface, this can be tested from the github page of the project. The app supports initial coding and themes generation, as well as visualisation. It offers OpenAI models as well as Azure. Albeit the notebook interface is not as user-friendly as a polished interface as running the analysis in cells may possibly be intimidating for users without programming experience.

Gao et al. (2025) proposed an application called *MindCoder*[5]. This is described in a pre-print article, and an online demo is available for some simple testing. The tool supports what the authors call a five steps workflow. In particular, the tool does a chunking of the data into similar themes using an LLMs, this is then followed by an automated sorting and coding of the grouped chunks. Codes are then aggregated themes and visualised.

A further app available on GitHub[6] is called 'l*lm_qualitative_data_analysis*', developed by Rudloff. This is not accompanied by a publication. From the GitHub repository, we can learn that the app creates summaries of the data and then the coding and themes generation is based on the summaries. The app uses streamlit as interface and langchain for creating LLMs applications. *QualiGPT*[7] by Zhang et al. (2023) is described as an easy-to-use tool for qualitative coding, and this implies the app only covers the coding step of qualitative analysis. The interface is based on the PyQt5 library/framework, the app can take textual data as input from Word documents or spreadsheets. PyQt5 is a framework better suited for Desktop applications but as such the UI components may have a marginally steeper learning curve than a GUI framework for websites.

*TRACER* (Transcript Analysis and Concept Extraction Resource) has been described in a paper by Sabbaghan (2024). It is a simple tool that in essence allows to upload a text and to extract initial codes, even though in the app interface these are called 'themes'. The app uses streamlit and langchain, and although it appears easy to use, it is modest as it does not cover a full analysis.

---

[2] See: https://github.com/ccbogel/QualCoder
[3] See: https://guides.temple.edu/qda/qualcoder
[4] See: https://github.com/PerttuHamalainen/LLMCode
[5] See: https://mindcoder.ai/
[6] See https://github.com/Gamma-Software/llm_qualitative_data_analysis

[7] See: https://github.com/KindOPSTAR/QualiGPT

# TALLMesh

Thematic Analysis with Large Language Models (TALLMesh) is the app we have developed for performing Thematic Analysis leveraging the capabilities of LLMs to manipulate text. The idea for the app came after one of our early publications on the topic. Indeed, once we put our first paper in pre-print, people started contacting us with interests for the method but also saying they did not have technical/programming skills to reproduce our work. From these early interactions with researchers showing interest, but lacking programming skills, came the realisation that having a graphical user interface could facilitate the adoption of the method. We thought that providing an easy and possibly intuitive GUI would bridge the gap between LLMs, python, prompt-engineering and practical, user-friendly tools for qualitative research.

For TALLMesh we decided to use a python framework known as streamlit[8], a software package that allows to easily build interactive web apps with python. Although this framework is for building websites, using a browser such as google chrome it is possible to run a website locally as if it was a standalone app. Streamlit also is a popular choice for developing apps using LLMs (see e.g. Thippeswamy et al., 20024; Jay, 2024).

The current version of the app TALLMesh has been developed in 3 different phases. In the first phase the pre-existing python scripts from De Paoli (2024a, 2024b) and De Paoli and Mathis (2024) were adapted to be used with streamlit. As the TA process (described earlier) is done in phases we used the multipage[9] option of streamlit with which we created a separate page for core phases of TA, including in particular initial coding, generation of themes and revisions of themes. Moreover, as the process we propose also requires a reduction of the codebook from duplicates, a further page was included for that. At this point the app was working with OpenAI models with credentials hardcoded in the scripts. In a second phase we made significant improvements to the quality of the code and worked in particular on keeping track of quotes for duplicate codes, included the option for adding API credentials (whilst removing them as hardcoded), and added a new page for creating and managing 'projects' with an easy process for uploading the data to be analysed. In the third phase we polished the interface and ran user testing to improve its usability. We also added new features such as visualisations, instructions and a prompt editor.

## Home Page of TALLMesh

---

[8] See: https://github.com/streamlit/streamlit
[9] See: https://docs.streamlit.io/develop/concepts/multipage-apps

The home page of TALLMesh (Figure 2) presents the menu of available features in the sidebar. This leverages the streamlit organisation of a webpage via the command st.sidebar, which divides the page in two main sections. Each of the features offered in the sidebar is in fact an independent (web)page, shown using the multipage feature of streamlit. The features of TALLMesh includes most importantly the 'Project Set-Up', and then the 'Initial coding' and 'Finding themes' step of TA (which allows to cover the phases 2 to 4 of Braun and Clarke's approach), in addition there is the 'Reduction of codes' step which we introduced to deal with saturation and avoid redundancies. The main feature will be presented in turn later. The home page also includes a list of the papers upon which the app is based, which provides references to users to learn more about the underpinning research. All the pages of TALLMesh follow this same organisation with the sidebar (and list of features) remaining constant and the main section of the page offering the features.

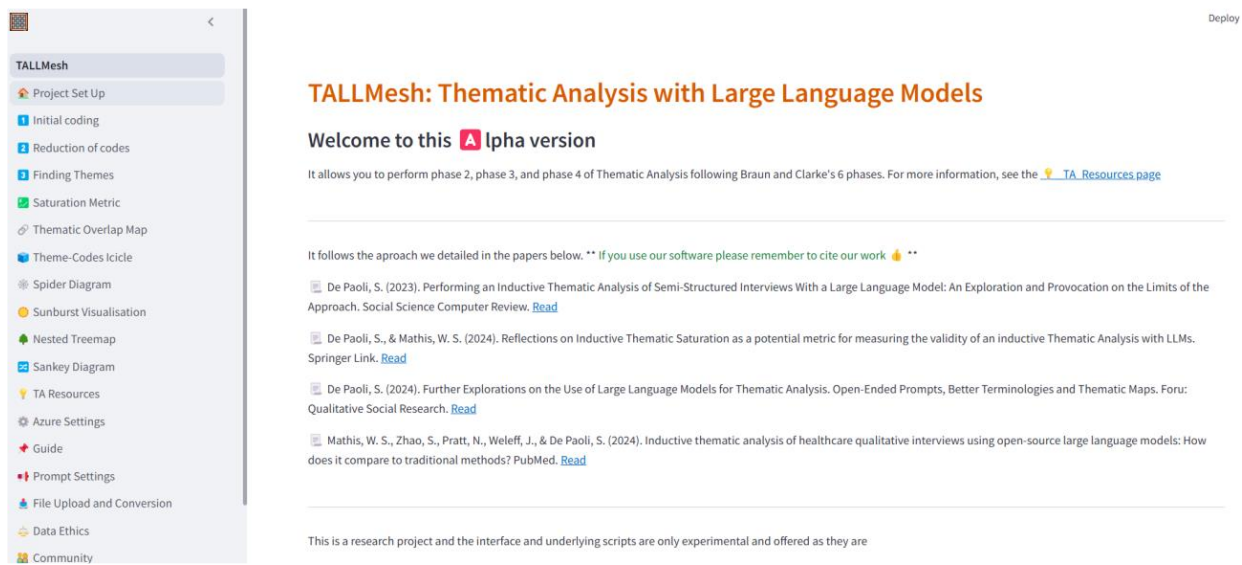

Figure 2. Home page of TALLMesh with main features listed on the sidebar

# Set-up and Guidance

## Guidance to users

The app offers guidance to users in dedicated pages (Figure 3) as well as within each page. A 'Getting started' page provides general guidance about the app, including describing the key features, and step-by-step instructions for setting up a project and running the whole analysis. A second page offers resources on TA, including presenting each of the phases of the method as defined by Braun and Clarke (2006) and discussing and comparing our own application of their approach into the app. This pretty much follows the process we described in our methodological papers.

## Welcome to TALLMesh: Thematic Analysis with Large Language Models

This application is designed to assist researchers and students in conducting thematic analysis using large language models (LLMs). Our approach is inspired by the widely-used Braun and Clarke method, adapted for the capabilities of modern AI.

TALLMesh aims to streamline the process of qualitative data analysis, offering a novel way to interact with your research data. While it leverages the power of LLMs, it's important to note that this tool is meant to augment, not replace, the researcher's critical thinking and interpretive skills.

Key features of TALLMesh include:

- Automated initial coding
- Assistance in reducing and refining codes
- Support in identifying and defining themes
- Integration with popular LLM providers

As you use this tool, remember that the AI is a collaborator in your research process. Always critically evaluate the outputs and use your expertise to guide the analysis.

> 💡 **Note for researchers:** While TALLMesh can significantly speed up parts of the thematic analysis process, it's crucial to maintain rigorous oversight. The tool's suggestions should be thoroughly reviewed and adjusted based on your research context and objectives.

### Getting Started

High level overview:

1. Set up your project and upload your data
2. Conduct initial coding phase
3. Reduction of duplicate/highly similar codes
4. Generate themes from reduced codes
5. Finalise theme book
6. Metrics and visualisations

---

**Phase 1 - Familiarizing Yourself with Your Data**
- Braun and Clarke's Approach
- Our App's Approach

**Phase 2 - Generating Initial Codes**
- Braun and Clarke's Approach
- Our App's Approach
- Additional Step: Reduction of Initial Coding

**Phase 3 - Searching for Themes**
- Braun and Clarke's Approach
- Our App's Approach

**Phase 4 - Reviewing Themes**
- Braun and Clarke's Approach
- Our App's Approach

**Phase 5 - Defining and Naming Themes**
- Braun and Clarke's Approach
- Our App's Approach

Figure 3. Part of the Getting started page (left) and of the TA Resources page

Moreover, each of the feature pages where the analysis is performed includes at the top additional tailored guidance and instructions, specifical for the action. For example, Figure 4 shows the instructions available for the Initial Coding phase. The instruction can be opened or closed using the arrow on the list element of the interface.

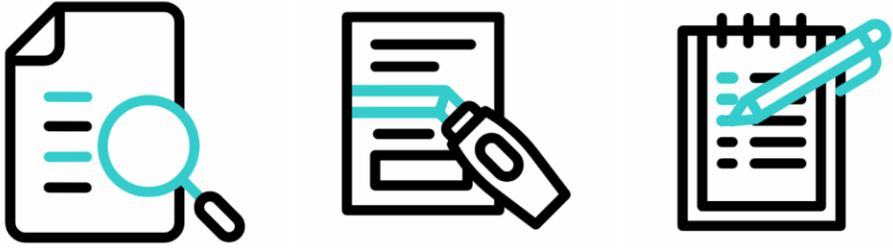

Figure 4. Example of in-page guidance

## Data Preparation

The current version of TALLMesh requires initial data to be text files (.txt extension). The user could create these files outwit the app of course, for example by saving interview transcripts from e.g. docx or .odt extension to .txt (using Unicode utf-8 encoding) using their word processing software (e.g. MS Word, Libreoffice). This is probably the easiest approach to get the initial data ready for processing. However, data formatting is critical to avoid errors generating from the python scripts, therefore one the features offered by the app allows to transform pdf files and .docx files into plain text (.txt Unicode utf-8 encoding). This feature is experimental, but is given as an additional option to prepare the starting dataset

Figure 5. Feature for converting files to plain text for processing

## Project management

For performing any analysis with TALLMesh a key substantive feature is the creation of a project. A project amounts to the analysis of one dataset, it contains the data and all the analysis results, including results from intermediate steps. The project set-up feature allows users to create a new/blank project by giving it a name (for example 'TA of interviews') and to associate with the project a set of data (e.g. a dataset of qualitative interviews in .txt) to it. Moreover, the project set-up creates the analysis folders where the intermediate products of the process are stored (e.g. a .csv file with the initial codes), as seen in Figure 6. These are local folders on the user computer, they are contained within the folder 'projects' inside the TALLMesh folder (e.g. C:\Users\Qual_Analysis\TALLMesh_multi_page\projects).

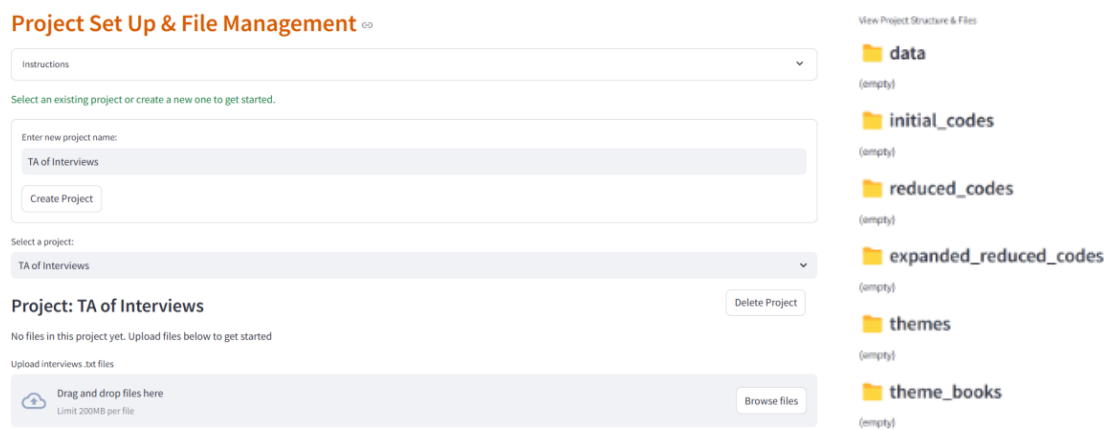

Figure 6. Project set-up and structure of the project as shown on the page

## API set-ups

For performing the key analysis steps the app offers currently the option to use OpenAI models, either directly using the OpenAI API, or via MS Azure using an institutional account (e.g. the MS Azure environment of the University of the researcher using the app). Both options are the same in terms of models, however the MS Azure institutional environment supports data processing in compliance with data protection and EU GDPR regulations. Albeit we always recommend researchers to check their compliance with their institution and ethics committee. Some aspects of data ethics are also presented in the dedicated page 'Data Ethics'.

For using the LLMs users need to possess their own credentials. For the general OpenAI model users need to obtain a secret API key, from their individual OpenAI account. To use the API the personal account also needs to have some credit for the processing of the tokens, purchased directly from OpenAI. Once the user has obtained the secret key this can simply be added in the API Key Management form which is located at the bottom of the homepage sidebar.

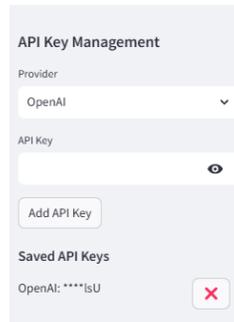

Figure 7. API Key Management form for OpenAI models

For using the LLMs within the Azure environment, users need to obtain a separate secret key from their own organisation. However additional details are required, including the endpoint of the Azure API and the name of the deployment. TALLMesh has a separate page that allows users to enter the Azure details and manage the different model deployments.

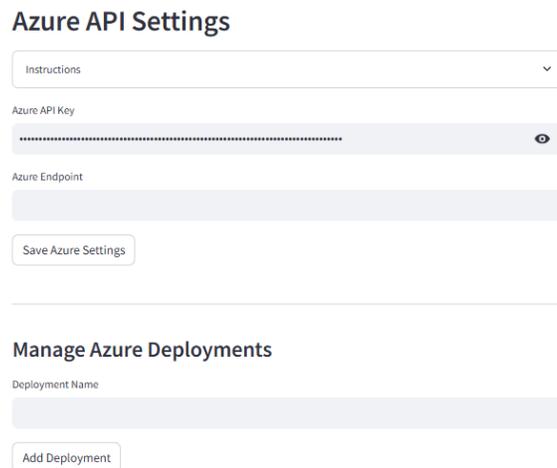

Figure 8. API set up and management for Azure

# Performing Thematic Analysis with TALLMesh

## Initial coding

Following the set-up of a project and of the API credentials, users can begin the Thematic Analysis. The first step is the initial coding. This feature/page of TALLMesh includes three main components:

1) The selection of the project on which the analysis is performed
2) The selection of the LLM to use and the prompt.

3) The output of the analysis

On landing on the Initial Coding page users will select the project upon which to perform the analysis from a drop-down menu, as seen in Figure 9. Once a project is selected the data files are loaded from the local folder data and are visible to the user. Each data file has a tick-box, this allows users to select and deselect the file. In this way the user can decide to perform the initial coding only on some files for example (Figure 10).

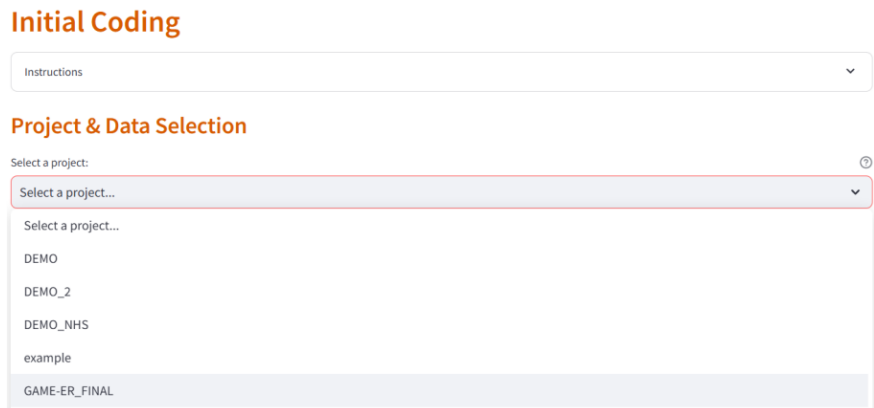

Figure 9. Project selection, Initial Coding

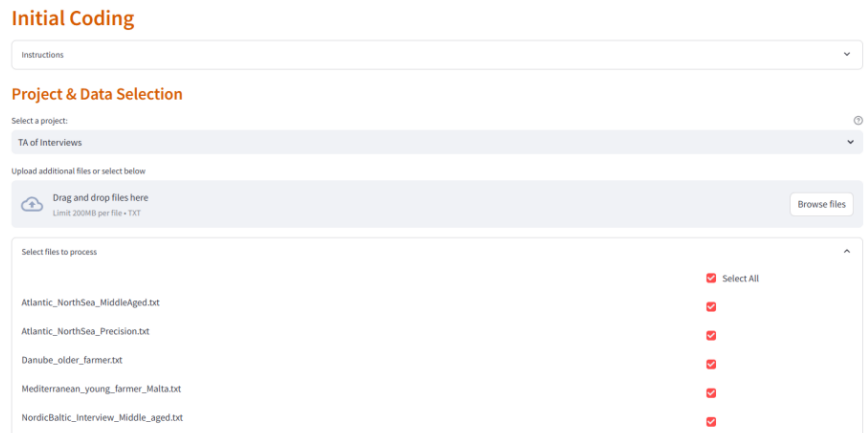

Figure 10. Data files selection for analysis

Following the project selection, users are required to set the LLM operation. First, they need to select the LLM to use for the analysis. Here again there is a drop-down menu for the selection of the models. The OpenAI models GPT-4, GPT-4o and GPT4o-mini are offered by default. The drop-down menu will also show any of those Azure models that were added by the user. In addition to the model, the user is allowed to select a prompt for performing the initial coding, from a drop-down menu. A few Preset prompts are available. The Preset prompts are tested and working for initial coding, based on our own research work. However, users can also define their own personalised prompt in a separate page of the app (see section Custom Prompts for this), and these user prompts will also be available from the menu. In any case, any of the prompts (including the Preset) can be directly modified by the user on this page. Lastly, the LLM user can set the Temperature and Top_P parameters to increase (or decrease) the LLM creativity. When all the parameters are at zero there is no creativity, and the response of the model is fairly deterministic. Increasing e.g. Temperature will allow the LLM to be more creative, however at the expense of reproducibility. After the LLM setting is completed, there is a button 'Process' which initiates the initial coding for the files, each done independently one at a time.

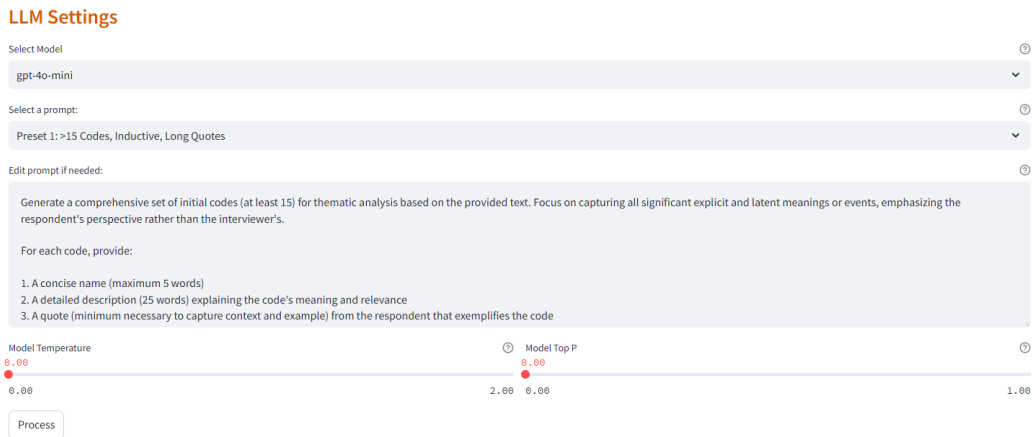

Figure 11. LLM setting section of the Initial Coding page

Figure 12 shows the output of the initial coding for one interview. This is returned to the user as a table, which is in essence a .csv file (one .csv will be generated for each file analysed). The .csv contains three columns: the initial code, a description of the code and a quote extracted from the file (size and length of these are set in the prompt). The .csv can also be downloaded by the user. However, all the .csv files are automatically and locally saved by TALLMesh in the initial_codes folder of the project and are available there for the users.

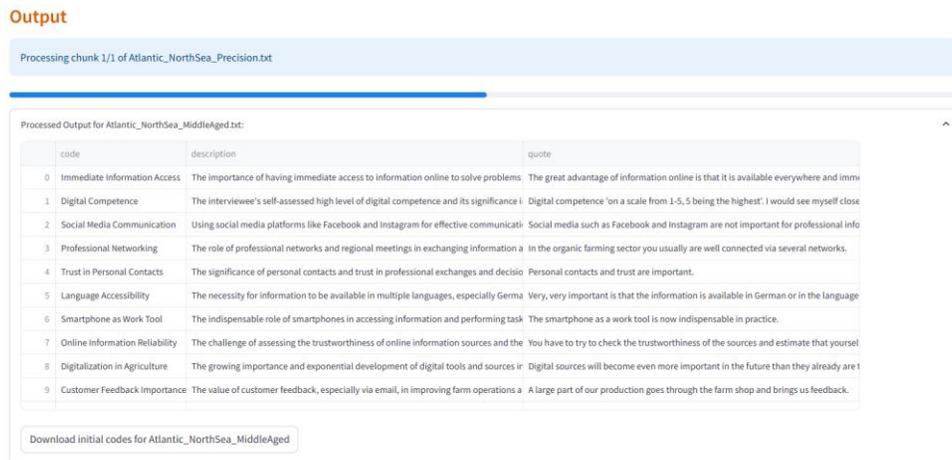

Figure 12. .csv output with initial coding for one file (one interview in this case)

## Reduction of codes

The reduction of codes is a step in the analysis we introduced to perform TA with LLMs. This has been described elsewhere in detail in our previous work. In simple terms, our approach entails producing initial codes for the files (e.g. interviews) independently one at a time. That is each file is coded one at a time independently from each other. This results in possible duplication of initial codes across the whole dataset. For arriving at a set of unique codes (without duplicates) - what we call the **unique codebook** – we have devised specific procedures (see De Paoli, 2024a, 2024b and De Paoli and Mathis, 20224), which are used by TALLMesh.

Similar to the initial coding users can select the project from the drop-down menu. This then will show the list of initial codes for each of the files processed/generated in the initial coding phase.

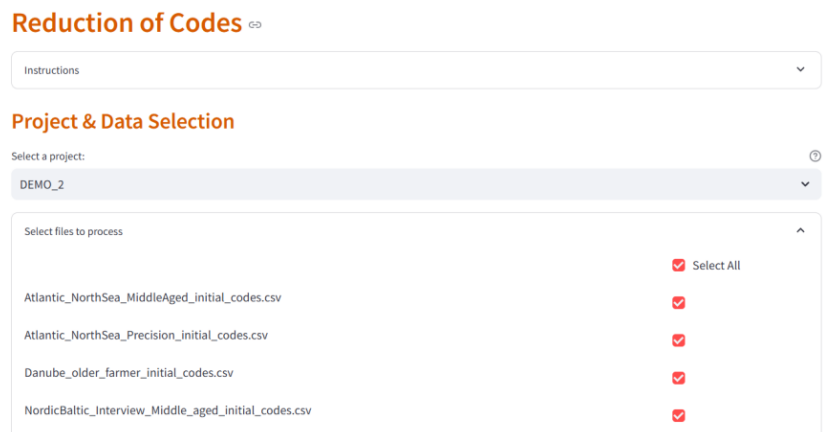

Figure 13. Reduction of Codes, project selection and initial codes files

The reduction of codes also has the LLM settings section, where users can select the model and the prompt. To note is that this step of the analysis operates a comparison of each code with the entire list of unique codes (the unique codebook), producing either a true result (the code already exists) or false the code is unique. When the code already exists, quotes are brought together, and the code name and description are updated. Readers can refer to De Paoli and Mathis (2024) for the underlying process and the definitions of unique and total codebooks.

One important aspect of the reduction of codes page is however the component following the LLM settings (Figure 14). Users can decide if quotes should be included in the comparison, and if they want the LLM also to generate and explanation for why codes were merged because identical of very similar.

The reduction of codes can be done currently in two modes (Figure 14), automatic and incremental. Automatic processes all the initial codes files together. Incremental allows users to process the initial codes files one at a time. As the reduction process takes some time to perform, automatic is recommended for small dataset (e.g. 10-15 interviews), whereas for larger datasets incremental is recommended. Moreover, incremental allows the analyst to keep better control of the process.

Figure 14. Options for the initial code reduction

The output of this process is a .csv file with the unique initial codes (the unique codebook). Moreover, the result of this process also allows the analyst to calculate what we have defined the Initial Thematic Saturation (ITS), given by the ratio of unique and total codes. This is a measure of saturation which relates the number of unique codes and the number of total codes (including duplicates). This metric is described in De Paoli and Mathis (2024). TALLMesh offers this as a feature/page called Saturation Metric, the feature provides a calculation of the metric and shows the graphical distribution of the unique and total codebooks (see Figure 15).

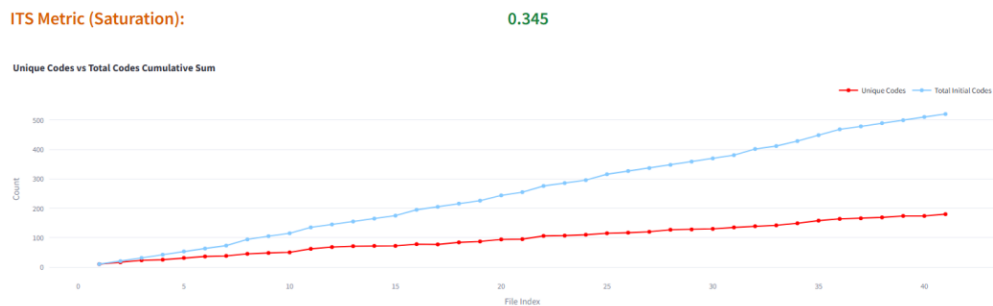

Figure 15. Saturation metric and graph of the codebooks

## Themes Generation

For the generation of Themes, the logic is similar to the one described earlier. First there is the selection of a project. This will then show the available unique codebooks. In case the reduction step was done automatically there will be only one unique codebook available, which will be used for generating the themes. If the reduction was done incrementally the unique codebook produced after each step is presented, and for generating themes users need to select only the last codebook in the list, as shown in Figure 16 (where only the last codebook is selected).

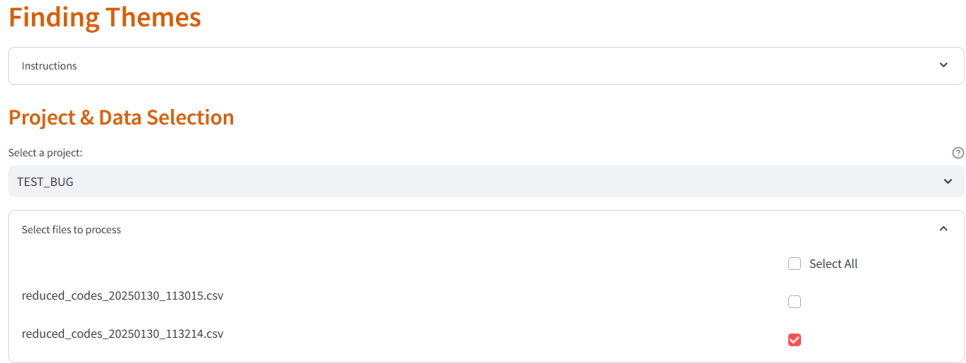

Figure 16. Available unique codebooks for a project, and selection of last one from incremental

Similarly to e.g. initial coding, the users then need to specify the LLM, select a prompt from those available (including those generated separately by the user) and set the model 'creativity'

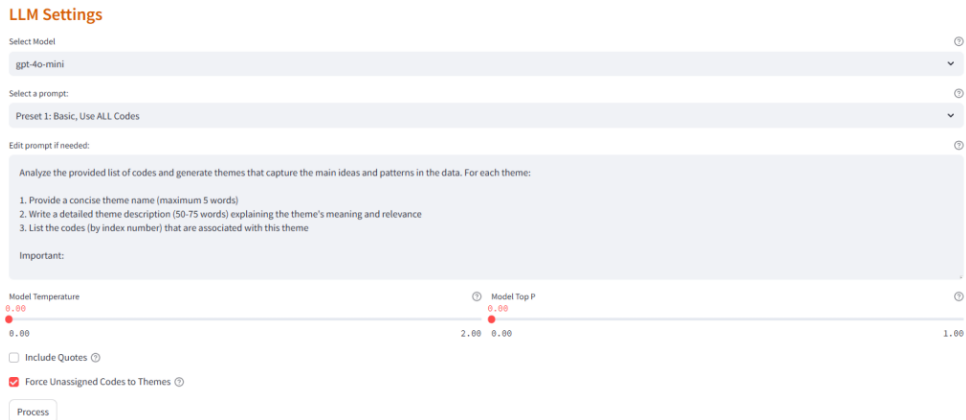

Figure 17. LLM settings and unassigned codes option

Before clicking on processing there are additional important options. The user can decide whether to include or not quotes in the generation of themes. We generally do not recommend selecting this option, and in this case the sorting and aggregation of codes into themes is done using the code label and its description. Lastly there is an option to force unassigned codes to themes. Effectively, this second option operates a two-step aggregation of codes. First the LLM assigns initial codes to themes, this may result in several codes remaining outwit the aggregation due to various reasons, including attention problems of the model. Second, the codes not assigned with the first pass are reconsidered are assigned to themes. This will leave out only codes which are entirely not compatible with any of the themes.

After the user clicks on the 'Process' button, the LLM operate the creation of themes. These will appear on the interface as drop-down menus, one for each theme generated. The .csv with the themes is also automatically saved in the 'Themes' folder on the user computer.

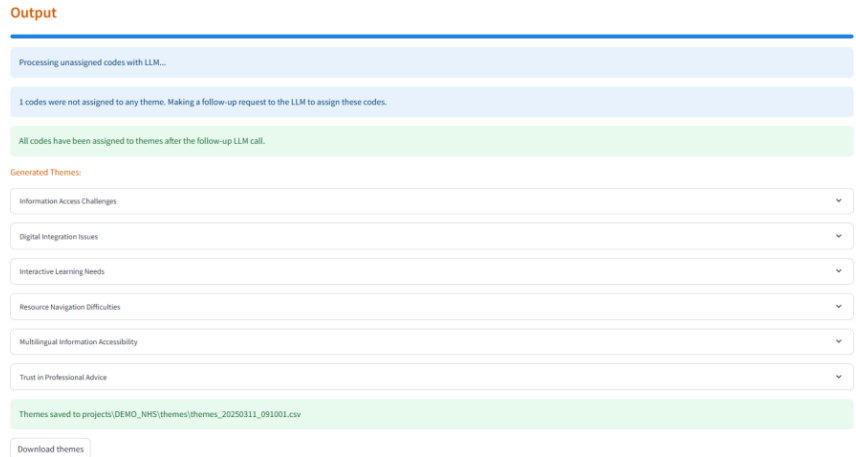

Figure 18. Example of themes generated by TALLMesh

Each drop-down menu can be opened, and the user will be able to see the full theme, including its description and the underlying codes aggregated within it, as shown in the example of Figure 19.

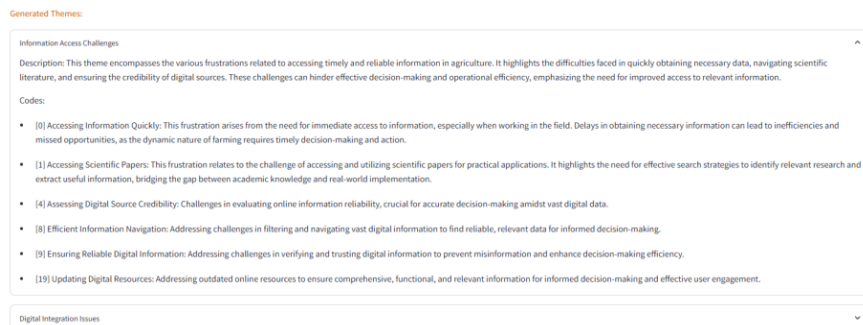

Figure 19. Example of one theme with underlying codes

## Visualisations

TALLMesh also offers visualisation features, providing intuitive and accessible ways to explore and interpret the outputs from the TA phases. Current visualisations are each an independent page of the app that can be accessed from the sidebar. The following visualisations are available: Thematic Overlap, Theme-Codes Icicle, Spider Diagram, Sunburst, Nested Treemap, and Sankey Diagram. We will briefly present the sunburst as an example. Like for the previous pages, users can select the project to visualise from a drop-down menu. Once a project is selected the Advanced Settings menu will open, which allows users to select what to visualise, including filtering based on themes and specific codes, as seen in Figure 20.

Figure 20. Sunburst selection and settings

Figure 21 shows a sunburst including themes and reduced/unique codes. Users can click on a specific theme also to visualise just that (left of Figure 21).

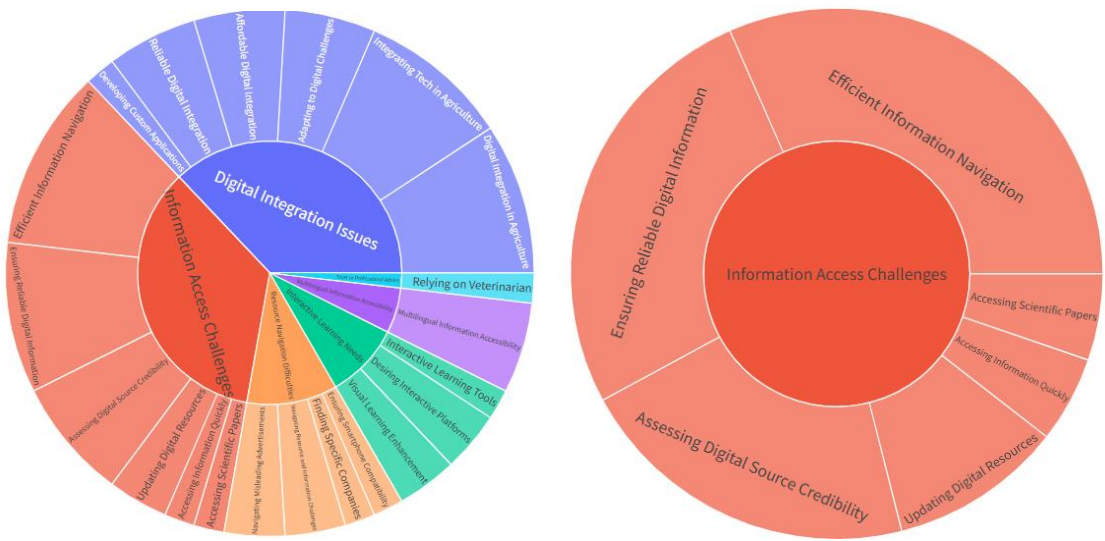

Figure 21. Example of sunburst visualisation of a set of Themes and one specific Theme.

## Custom prompts

One of the features of TALLMesh is allowing users to create their own custom prompts, as previously discussed in relation to initial coding and finding themes. This is a standalone page. The user can select for which step the custom prompt is prepared (initial coding, finding themes, reduction), then can name the new prompt and edit the new prompt text in the form (see Figure 22). One important aspect however is that the formatting section of the prompt must remain standard for the analysis to work properly. The json formatting is already available in the edit form text. Once the custom prompt is prepared users can simply click on the Add Custom Prompt button and then the prompt will be available in the respective step of the analysis in the app.

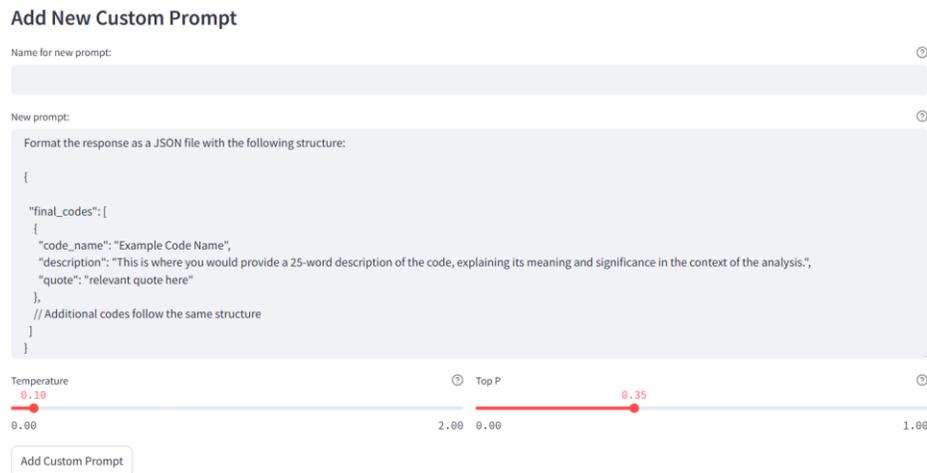

Figure 22. Custom prompt set-up

The Custom Prompt feature allows users to tailor their analysis. For example to specific research questions or specific requirements, such as a fixed number of themes or by providing clear definitions about what is sought from the data.

## User Testing

To improve the flow and clarity of the interface, we conducted user testing at our university in the form of a 'think-aloud' protocol. Think-aloud is a technique which invites participants to perform some tasks with an app or solution whilst verbalising what they are doing. The protocol also entails minimal or no intervention from researcher performing the user testing. We assigned tasks to users ranging from adding an API key, create a project, perform steps of the analysis, delete documents, visualise data and so forth

A Miro board was then used as an analysis tool to organise our observations, highlight interface problems, and track potential solutions. We mapped participant feedback, annotated the screenshots of interface issues together with key quotes from transcriptions, and documented errors or issues, at the same time identifying possible actions to address them. Figure 23. Shows the full Miro board with our analysis (on the left) and two specific examples (on the right)

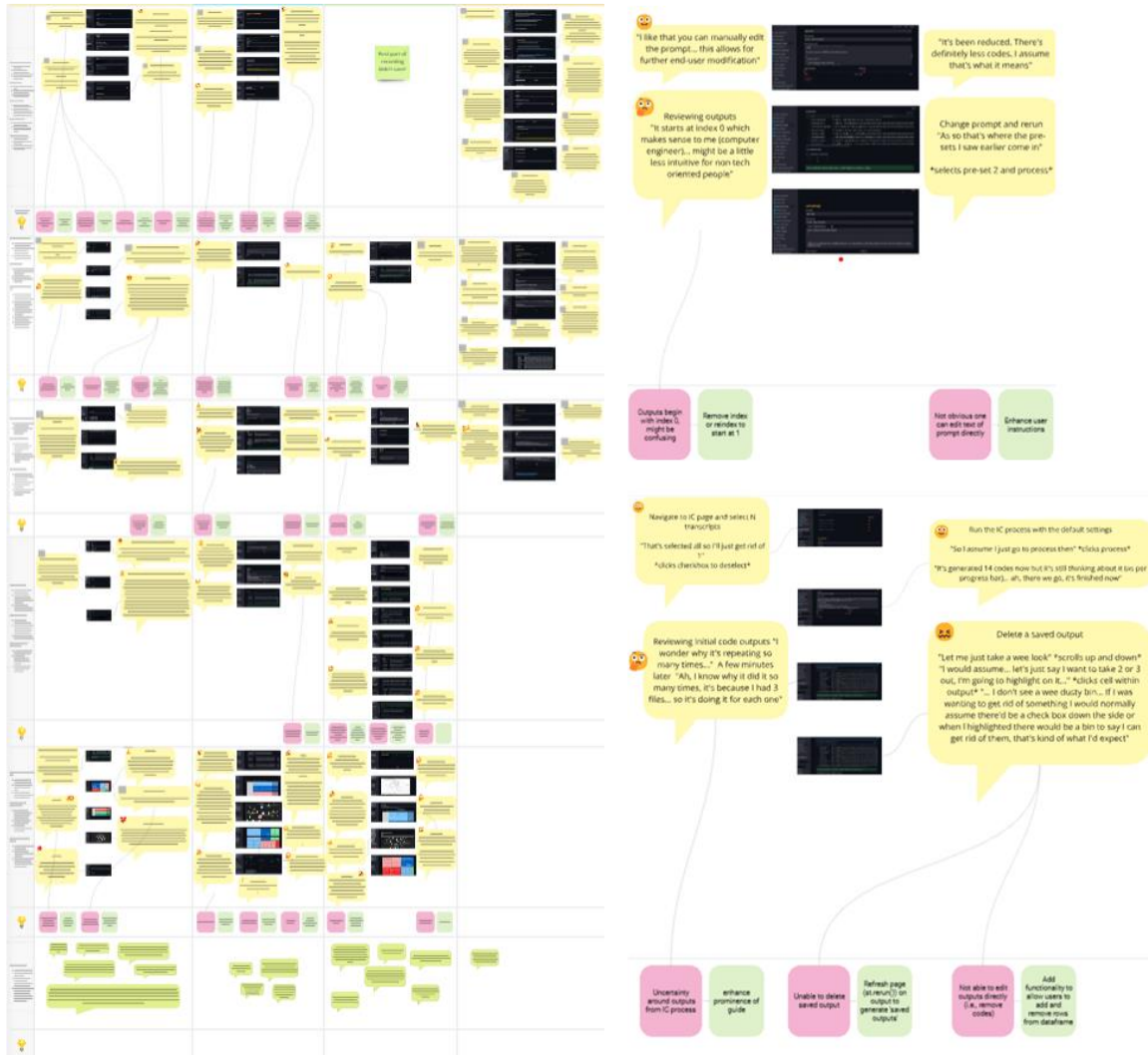

Figure 23. User testing analysis

The user testing allowed us to identify and resolve several interface issues. This resulted in an overall improvement of the usability and clarity of the GUI. Key enhancements included:

- Labels - Occasionally participants struggled with unclear labels (for example on buttons), making it difficult to understand the specific functionality or action required. Based on the collected feedback, we were able to better refine button names and some menu items.
- Feedback - Users sometimes were unsure whether an action was successful during processing, or whether they could do something (e.g. like edit a Preset prompt). To address this, we implemented progress indicators, and clear and simple success/error messages.
- Bug fixes and errors - Some participants, whilst performing the assigned tasks encountered unexpected errors. These included missing libraries, unresponsive elements, and incorrect outputs. With these observations and through detailed debugging, we fixed these issues and improved error messages. This should provide better guidance to user in case of errors.

# Discussion and Future Work

This paper has described the app for performing Thematic Analysis with LLMs called TALLMesh. The overall objective of TALLMesh is making TA performed with LLMs more accessible. One of the primary contributions of TALLMesh is indeed its ability to facilitate TA with LLMs without requiring extensive programming knowledge. The use of the streamlit framework allows to build intuitive interfaces that are easy to use. This is particularly important for researchers in fields where coding skills may not be prevalent, such as social sciences or humanities. TALLMesh offers a user-friendly interface based on streamlit and automates key phases of TA with python, LLMs and use of APIs. It also offers additional features such as visualiations and custom prompting, allowing users more control over the results and the analysis.

Despite these advances, there are still areas where we feel TALLMesh can be improved. One key area is the creation of more comprehensive documentation and support materials for TALLMesh, such as video tutorials, and step-by-step workflow instructions. While the user testing we conducted helped us to identify and address a number of issues related to the interface, we are aware that some users may still require additional guidance or training to get the most out of the tool. By providing clear, concise documentation and tutorials, we can help to ensure that TALLMesh can become accessible to an even broader range of researchers. We are also working actively to disseminate the app to relevant scientific communities, for example in March 2025 we presented the app to circa 100 scholars from the Early Career Research Network of the British Academy[10], and are looking to repeat this event in late 2025. We are also seeking to actively build a discord online community where researchers can discuss the app, and the application of LLMs to qualitative analysis more generally (discord server name: *TALLMesh - Thematic analysis with LLMs and more*).

On a more substantive level, we are aware that new processes are being worked out for performing the thematic analysis steps. For instance, in a recent pre-print paper we described a novel process for the codebook reduction and for the creation of a unique codebook (see De Paoli and Mathis, 2025). This novel technique appears more reliable than the current one used within TALLMesh and it will need to be incorporated into the app soon. We are also looking at the possibility of offering Open-Source LLMs for performing the analysis and are exploring some options for implementing code that can support this. TALLMesh is also one of the experimental pilots of a recently funded Horizon project called GRAPHIA[11] (Knowledge Graphs, AI Services and Next Generation Instrumentation for Research and Development in Social Sciences and Humanities), where the app will be further enhanced and more widely tested with the social sciences and humanities communities.

---

[10] See: https://www.tickettailor.com/events/earlycareerresearchernetwork/1508940
[11] See project page: https://cordis.europa.eu/project/id/101188018

TALLMesh represents a positive step for making TA with LLMs accessible to non-programmers. Our future work will focus on building upon this foundation, expanding the app capabilities, and providing additional support for users.